\documentstyle[aps,pre,amssymb,rotating,epsfig,amstex]{revtex}

\def\bra{\langle}
\def\ket{\rangle}
\def\mum{\mu{\rm m}}
\DeclareMathSymbol{\BBB}{\mathbin}{AMSb}{'102}
\DeclareMathSymbol{\KKK}{\mathbin}{AMSb}{'113}
\DeclareMathSymbol{\MMM}{\mathbin}{AMSb}{'115}
\DeclareMathSymbol{\PPP}{\mathbin}{AMSb}{'120}
\DeclareMathSymbol{\SSS}{\mathbin}{AMSb}{'123}
\DeclareMathSymbol{\DDD}{\mathbin}{AMSb}{'104}
\DeclareMathSymbol{\ZZZ}{\mathbin}{AMSb}{'132}
\newcommand{\cali}{\hbox{\raisebox{.4ex}{$\chi$}}_\PPP}
\newcommand{\LiPa}{\hbox{\sffamily L}}
\newcommand{\PoSi}{\hbox{\sffamily P}}
\newcommand{\TwPo}{{\hbox{\sffamily S}}_2}
\newcommand{\LS}{{\sffamily LS }}
\newcommand{\PS}{{\sffamily PS }}

\def\figref#1{Figure \ref{fig:#1}}
\def\FIGDIR{./Figures}

\begin{document}
\bibliographystyle{prsty}

\title{Stochastic reconstruction of sandstones}

\author{C.~Manwar${\rm t}^1$, S.~Torquat${\rm o}^{2}$, and R.~Hilfe${\rm r}^{1,3}$}

\address{$\ ^1$ Institut f\"ur Computeranwendungen 1, Universit\"at
  Stuttgart, 70569 Stuttgart, Germany\\ 
  $\ ^2$ Department of Chemistry and 
  Princeton Materials Institute, Princeton University, Princeton, NJ
  08544, USA \\
  $\ ^3$ Institut f\"ur Physik, Universit\"at Mainz, 55099 Mainz,
  Germany} 
  
\date{\today}

\maketitle

\begin{abstract}
  A simulated annealing algorithm is employed to generate a stochastic
  model for a Berea and a Fontainebleau sandstone with prescribed
  two-point probability function, lineal path function, and ``pore
  size'' distribution function, respectively. We find that the
  temperature decrease of the annealing has to be rather quick to
  yield isotropic and percolating configurations. A comparison of
  simple morphological quantities indicates good agreement between the
  reconstructions and the original sandstones. Also, the mean survival
  time of a random walker in the pore space is reproduced with good
  accuracy. However, a more detailed investigation by means of local
  porosity theory shows that there may be significant differences of the geometrical
  connectivity between the reconstructed and the experimental samples.
\end{abstract}

\bigskip
\begin{tabbing}
  PACS: \= 61.43Gt \hspace*{2ex}\= (Powders, porous materials)
\end{tabbing}

\twocolumn
\clearpage
\section{Introduction}
The microstructure of porous media determines their macroscopic
physical properties \cite{B:sah95,hil96,tor91} such as conductivity,
elastic constants, relaxation times, permeabilities or thermal
properties. The relation between geometric microstructure and physical
properties is a fundamental open problem whose solution is important
to many applications ranging from geophysics to polymer physics and
material science. For geometric modeling of porous media it is
therefore important to characterize microstructures quantitatively and
to construct models using given geometric characteristics.

The reconstruction of random media with given stochastic properties is
also of great interest for a variety of other reasons. (i) Digitized
3D geometries of real sandstones are difficult to obtain. Therefore,
the reconstruction provides a method for easily generating detailed
geometries as needed e.g. in numerical calculations of macroscopic
material parameters like those mentioned above. (ii) The
reconstruction of 3D samples from 2D data. (iii) Any calculation of
macroscopic quantities of random media needs a set of stochastic
functions which describes the geometry. The reconstruction can help to
decide which functions one should use. The present work will focus
mainly on the latter question, the so-called inverse problem.

Recently, a simulated annealing algorithm for the reconstruction of
random porous media with predefined stochastic functions was proposed
\cite{YT98a}. This method was used for the reconstruction of two-phase
porous media, i.e. sandstones, with given two-point probability
function and lineal-path function, which were measured from real
sandstones \cite{YT98b,BMHBO99}. In the present article, we will
continue this work.  An advantage of the simulated annealing is that
it allows the reconstruction of a variety of different stochastic
functions, with the available CPU time being the only limit on the use
of a set of functions. In Refs. \cite{YT98b,BMHBO99}, the two-point probability
function and the lineal path function were used. Here we will extend
the investigations to the reconstruction of the so called ``pore
size'' distribution \cite{B:SCH74,TA91}, which is
not the usual quantity obtained from mercury porosimetry.
We will study a Berea sandstone and  a Fontainebleau sandstone.

A comparison of the reconstructions with the original sandstones shows
to which extent the characteristics of the original geometry are
reconstructed. A salient feature of the original sandstone is the high
degree of connectivity of the pore space.  Hence, one important
criterion for judging a stochastic reconstruction method is its
ability to reproduce the connectivity of the original sandstone.
It is shown that Berea reconstruction does a better job
of capturing connectivity than does the Fontainebleau reconstruction,
although neither captures connectivity particularly well.
Another important criterion is the ability of the reconstruction procedure
to reproduce the macroscopic properties of the original materials.
We show that the mean survival times of our reconstructions of both sandstones 
agree well with mean survival times of the original sandstones.

The paper is organized as follows. In section II we describe the
reconstruction algorithm. In section III we introduce the
quantities we use in the reconstruction process and for characterizing
the sandstones. Because detailed discussions can be found in the
references, we will focus mainly on practical aspects and details of
the implementation. In section IV we present the results for
reconstructions of a Berea and a Fontainebleau sandstone, respectively.

\section{The reconstruction method}
A two-phase porous medium consists of a pore or void phase $\PPP$ and
a matrix or rock phase $\MMM$. Its microgeometry is described in detail by
the characteristic function
\begin{equation} \label{eq:I}
  \cali(\vec x) = \left\{ 
  \begin{array}{ll}
     0 & {\quad \rm for \quad} \vec x \in \MMM \\
     1 & {\quad \rm for \quad} \vec x \in \PPP 
  \end{array}
  \right.
  .
\end{equation}
For a discretized sample $\vec x=(j_1a,j_2a,j_3a)$ is the position
vector of a cubic grid with $j_i=0,\dots,M_i-1$, lattice constant $a$
and size $M_1\times M_2\times M_3$. The total number of grid points is
given by $N=M_1M_2M_3$. In the following we will refer to a grid point
also as a voxel.

The porosity $\phi$, the probability of finding a point in pore space,
is given by
\begin{equation}
  \phi = \bra \cali(\vec x)\ket
  .
\end{equation}
With the assumption of a homogeneous, stationary, and ergodic
stochastic porous medium the angular brackets denote a volume average.

The reconstruction is carried out by means of a simulated annealing
method \cite{YT98a} with target or ``energy''-function defined as
\begin{equation}\label{eq:energy}
  E_t(f_t) = \sum_{\vec x} \left|f_t(\vec x) - f_{\rm ref}(\vec x)
  \right|^2 .
\end{equation}
The function $f_{\rm ref}$ is the stochastic function to be
reconstructed, whereas $f_t$ is the actual value of this function
measured at iteration step $t$. For the reconstruction of more than one
function the energy $E_t$ at the iteration step $t$ is given by
$E_t=\sum_k E_t(f^{(k)}_t)$ where the index $k$ numbers the different
functions to be reconstructed. 

Starting from a random configuration with porosity $\phi$ two voxels
of different phase are exchanged at each iteration step. Thus, the
porosity $\phi$ remains constant during the reconstruction
process. The new configuration is accepted with the probability given
by the Metropolis rule
\begin{equation}\label{eq:prule}
  p = 
  \left\{\begin{array}{ll}
    1 & \qquad {\rm if} \quad  E_t \le E_{t-1} \\
    e^{\frac{E_{t-1}-E_t}{T}} & \qquad {\rm if} \quad E_t>E_{t-1}       
  \end{array}\right.
  ,
\end{equation}
where $T$ plays the role of a temperature. In the case of rejection,
the old configuration is restored. By decreasing the temperature $T$
configurations with minimal energy $E$, i.e. with minimal deviations
of the stochastic functions $f^{(k)}_t$ from their reference functions
$f^{(k)}_{\rm ref}$, are generated.  The process terminates after a
certain number of consecutive rejections. Here, the reconstruction was
finished after $10^5$ consecutive rejections.

\section{Measured quantities}

\subsection{The two-point probability function}

Using Equation (\ref{eq:I}) the two-point probability function is
defined as
\begin{equation}\label{eq:g}
  \TwPo(\vec x_1, \vec x_2) = \bra\cali(\vec x_1)\cali(\vec
  x_2)\ket
  .
\end{equation}
For a homogeneous and isotropic medium $\TwPo(\vec x_1,
\vec x_2) = \TwPo(r)$ with $r = |\vec x_1 -\vec x_2|$ holds. In this case
$\TwPo(r)$ can be evaluated without loss of information from the
intersection of the sample with a plane or even a line. To speed up
the numerical evaluation of $\TwPo$ it is therefore sufficient to sample
$\TwPo$ only in directions of the principal axis given by the unit vectors
$\vec e_i$ \cite{YT98a}.

Hence, $\TwPo$ is calculated by evaluating equation (\ref{eq:g}) for
every pair of voxels at $\vec x_1$ and $\vec x_2 = \vec x_1 + r\vec
e_i$ with $r=0,1,\dots,r_c$ where $r_c$ is a cut off value determined
by the system size or a multiple of the correlation length. During the
reconstruction only those terms in $\TwPo$ are updated that have changed due
to the exchange of voxels. On the sample boundaries we impose periodic
boundary conditions.

For reconstruction purposes this simplification, i.e. the
reconstruction of $\TwPo$ only in the direction of the three coordinate
axes, may create some problems. Because all other directions remain
unoptimized, $\TwPo$ measured in these directions may differ from the
reference function. If this happens the reconstructed sample is no
longer isotropic \cite{MH99,CT99}. Furthermore, the reconstruction of
$\TwPo$ along orthogonal lines only, reduces the three-dimensional
optimization problem effectively to the optimization of three
one-dimensional two-point probability functions. This may reduce the number of
conditions that the two-point probability function has to fulfill to be
realizable \cite{tor99}, and may lead one to conclude incorrectly that
the reconstruction is realizable in the space dimension of interest.
However, the problem of realizability does not apply to the
reconstructions presented here, because the two-point probability functions used
as reference functions are measured from digitized three-dimensional
images of real sandstones. Nevertheless, the reconstructions have to be
checked for their isotropy.

The isotropy of the reconstructions can be improved e.g. by a full
reconstruction of the two-point probability function using Fourier transform
techniques or by rotating the sample during the reconstruction. This
was done for two-dimensional reconstructions in \cite{CT99}. However,
for three-dimensional reconstructions this leads to prohibitive
increase of computation time.

For the specific surface $s$, i.e. the surface per unit volume of the
interface between pore space and matrix space, it is known that
\cite{DAB57}
\begin{equation}\label{eq:ss}
  s = -4\left.\frac{\partial \TwPo(r)}{\partial
  r}\right\vert_{r=0}
  \quad.
\end{equation}
Therefore, a reconstruction of $\TwPo$ implies that the specific surface
area of the reconstructions matches that of the reference
sample. Equation (\ref{eq:ss}) may also be used for the calculation of
the specific surface. Here, we use a different, numerically very
efficient method introduced in \cite{LOH99} for calculating $s$.

\subsection{Lineal path function and ``pore size'' distribution}

The lineal path function $\LiPa(r)$ is defined to be
the probability of finding a
line segment of length $r$ entirely in pore space
when the line segment is randomly thrown into
the porous medium \cite{Lu92}. Hence, for $r=0$,
$\LiPa(0)=\phi$ holds. The lineal path function 
is related to the linear contact
distribution introduced in mathematical stochastic geometry
\cite{del72,B:SKM87,lpnote}. 
The lineal path function is calculated as
follows: For a given pore voxel the lineal path $r$ in units of the
resolution $a$ is given as the number of pore voxels lying between the
given pore voxel and the nearest matrix voxel in direction $\vec e_i$.
Evaluating the line segments starting from each pore voxel in all
three coordinate directions $\vec e_i$ and counting the number $l(r)$
of line segments with length $r$, the lineal path function is given by
\begin{multline}\label{eq:lineal}
  \LiPa(r) = l(r) / \left( (M_1-r)M_2M_3 \right.\\ \left.+ M_1(M_2-r)M_3 + M_1M_2(M_3-r) \right)
\end{multline}
where we assume non periodic boundary conditions.

The lineal path function incorporates information about the
connectivity of the pore space. Of course it would be desirable to
reconstruct functions which provide a more complete information about
the geometric connectivity of the pore space, as for example the
cluster correlation function \cite{tor91} or the local percolation
probability \cite{hil96}, but currently the computation time for
evaluating these functions prevents their use in the above
reconstruction scheme.

The ``pore size'' distribution function $\PoSi(\delta)$ is defined
such that $\PoSi(\delta){\rm d}\delta$ is the probability that a
randomly chosen point in the pore space lies at a distance $[\delta,
\delta + {\rm d}\delta]$ from the nearest point on the interface
\cite{B:SCH74,TA91}.  It is related to the spherical contact
distribution \cite{del72,B:SKM87,pnote}.  The associated cumulative
distribution function
\begin{equation}
  \hbox{\sffamily F}(\delta_o)=\int_{\delta_o}^\infty \PoSi(\delta) {\rm d}\delta
\end{equation}
gives the fraction of the pore space which has a diameter larger
then $\delta_o$. Clearly 
\begin{equation}
  \PoSi(\delta) = -\frac{\partial \hbox{\sffamily F}}{\partial \delta}
\end{equation}
and
\begin{equation} \label{eq:pnod}
  \PoSi(0) = \frac{s}{\phi}
  .
\end{equation}  
The mean pore size is given as 
\begin{equation} \label{eq:delta}
  \bra \delta \ket = \int_0^\infty \delta \ \PoSi(\delta) {\rm d}\delta
   = \int_0^\infty \hbox{\sffamily F}(\delta) {\rm d}\delta
  .
\end{equation}
The quantity $\PoSi(\delta)$ arises in rigorous bounds on the mean
survival time \cite{TA91}.

We compute an approximation to $\PoSi(\delta)$ by choosing
a random point in pore space and measuring its distance $\delta$ to
the nearest point on the matrix-pore interface assuming periodic
boundary conditions. This process is repeated for several random
points in pore space. The ``pore size'' distribution is then obtained by
binning the distances $\delta$ and dividing by the number of random
placements in pore space. We emphasize that the random placements in
pore space are not necessarily grid points. The computation of
$\delta$ is only approximate because it requires a modeling of the
interface between pore and matrix space. Here, we assume that the
internal surface is given by the surface of the cubic voxels. This is
the same modeling which is used e.g. in a computation of the mean
survival time or in finite difference calculations of transport
properties. In general this may overestimate the specific surface area
appearing in Equation (\ref{eq:pnod}) by a factor of roughly $1.5$
\cite{CT95b}.

For use in the reconstruction we measured $\delta$ as the distance
between a pore voxel and the nearest matrix voxel. The resulting
function is not equal to $\PoSi(\delta)$ because $\delta$ can now only take
values of $\delta = \sqrt{i^2 + j^2 + k^2}$ with $i,j,k \in \ZZZ$.

\subsection{The total fraction of percolating cells}

The total fraction of percolating cells $p$ is a key measure in local
porosity theory \cite{hil92,hil96}. Local porosity theory measures the
fluctuations of morphological quantities, e.g. porosity, specific
surface, connectivity, in cubic subsamples of the total sample
\cite{hil00}. We will refer to such a cubic subsample with side length
$L$ as a measurement cell. Based on the scale-dependent morphological
quantities, the theory provides scale-dependent estimates for transport
parameters from a generalized effective medium theory \cite{WBH99,HWB00}.

Let $\mu(\phi,L){\rm d}\phi$ be the probability that a given
measurement cell of side length $L$ has a porosity in the interval
$[\phi,\phi+{\rm d}\phi]$. The probability density function $\mu$ is
called local porosity distribution. The probability that a given
measurement cell with porosity $\phi$ is percolating in all three
directions is the local percolation probability function
$\lambda(\phi,L)$. Here, percolating in all three directions means
that each face of the measurement cell is connected to the opposite
face with a path lying entirely in pore space. Using this, the total
fraction of percolating cells is given by
\begin{equation} \label{eq:lptp}
  p(L) = \int_0^1 \mu(\phi,L) \lambda(\phi,L) {\rm d}\phi
  .
\end{equation}
The total fraction of percolating cells $p(L)$ is the probability of
finding a measurement cell with side length $L$, which is percolating
in all three directions. Hence, $p$ is a measure for the geometrical
connectivity.

\subsection{The mean survival time}

Contrary to the previous quantities the mean survival time $\tau$ is
not a purely geometrical but a physical observable. The mean survival
time $\tau$ is the average life time of a random walker, which can
freely move in pore space, but gets instantly absorbed on contact with
the pore-matrix interface. It is a measure of a characteristic pore
size. The mean survival time $\tau$ is calculated using a first
passage cube (FPC) algorithm \cite{CT95b,TKC99}. The FPC algorithm
uses the fact that the mean time it takes for a certain type of
diffusive random walker starting at the center of a cube with side
length $2L$ to cross the surface of this cube is given by
\begin{equation}\label{eq:tau}
  \tau(L) \approx 0.22485L^2
  .
\end{equation}
Hence, it is not necessary to simulate the steps of the walker in
detail. Instead one determines the biggest cube centered around the
position of the walker, which is still entirely in pore space. The
walker then jumps on the surface of this FPC and a time given by
Equation (\ref{eq:tau}) is added to its life time. This procedure is
iterated until the walker touches the interface and gets absorbed. The
mean survival time is given by averaging over many walkers.

The probability with which the walker jumps to a certain point on the
surface of the FPC is described by a probability density function
$w(y,z)$ where $y$ and $z$ are the coordinates on the surface assuming
without loss of generality that $x=\pm L$. For an analytic expression
of $w(y,z)$ we refer to \cite{TKC99}.

\section{Results}
In this section we present results for reconstructions of a Berea
sandstone and a Fontainebleau sandstone. For both sandstones we
computed reconstructions with the lineal path function and the
two-point probability function (\LS reconstruction) as well as
reconstructions with the ``pore size'' distribution function and the
two-point probability function (\PS reconstruction). The Berea
sandstone and its reconstructions have dimension
$128\times128\times128$ and resolution $a=10\mum$. The porosity is
$\phi=0.1775$. The Fontainebleau sandstone has dimensions $299\times
300\times 300$ its reconstructions have dimension
$128\times128\times128$. The resolution is $a=7.5\mum$, the porosity
is $\phi=0.1355$. The reconstructed functions were calculated as
described above, i.e. we used periodic boundary conditions except for
the lineal path function. The two-point probability function
$\TwPo(r)$ was reconstructed in the interval $r=0,1,\dots,63$. The
annealing process terminated after $10^5$ subsequent rejections.

We performed 5 \LS reconstructions and 5 \PS reconstructions for both
the Berea and the Fontainebleau sandstone. Some of the results are
summarized in Table \ref{tb:berea} for the Berea and Table
\ref{tb:fntbl} for the Fontainebleau sandstone, respectively. The
values are averaged over 5 reconstructions.  The quantity 
$\tau D$ is the mean survival time multiplied by the 
diffusion coefficient $D$ for the random-walk process 
discussed above. The quantity $f_p$ gives
the fraction of pore voxels which belong to the percolating cluster.

\figref{slices} shows 2D slices of the original and the reconstructed
sandstones. The top row shows the original sandstones, the row in the
middle the \LS reconstructions, and the bottom row the \PS
reconstructions. The slices were taken from reconstructions with the
value of $p(L=60)$ close to the average values given in the
tables. All slices are chosen to have average porosity. In the case of
the Berea sandstone the reconstructions look similar to the original
sandstone while for the Fontainebleau sandstone the reconstructions
are clearly distinguishable from the original sandstone. The matrix of
the original Fontainebleau sandstone shows a granular structure where
single grains can be identified. The pores between these grains are
long and narrow. In the reconstructions no granular structure of the
matrix space is visible. The pores of the reconstructions are more
rounded in shape. For both sandstones the number of isolated pores is
significantly higher in the reconstructions. This is expressed by
$f_p$, the fraction of pore space belonging to the percolating
cluster. For the Berea $97.16\%$ of the pore space belongs to the
percolating cluster whereas for the reconstructions this fraction is
roughly $10\%$ smaller. For the Fontainebleau the difference is even
bigger. In the original sandstone $99.35\%$ of the pore space
percolates while for the reconstructions $f_p$ is approximately
$52\%$. Here we find that one \LS and one \PS reconstruction of the
Fontainebleau is not percolating in all three directions.

In the course of our work we used also a slowly decreasing step
function for the temperature $T$ to obtain an optimal match of the
reconstructed functions. Surprisingly, with a slow cooling schedule
the majority of the reconstructed configurations was not percolating
in all three directions. Furthermore, the reconstructed samples showed
a strong anisotropy with $E(\tilde\TwPo)$ of order $10^{-2}$ where
$\tilde \TwPo$ denotes the two-point probability function measured in
the directions $e_i+e_{j\ne i}$.

For the reconstructions presented here, we used a fast exponential
cooling schedule $T = \exp\left(\frac{t}{10^{5}}\right)$ where $t$ as
above denotes the iteration step. This cooling schedule took
approximately $30\cdot N$ iteration steps to complete a
reconstruction, whereas the slow cooling took more than $300\cdot N$
iterations steps. Using the fast cooling schedule all reconstructions
of the Berea sandstone are percolating in all three directions and
only one \LS and one \PS reconstruction of the Fontainebleau is not
percolating. Also with the fast cooling schedule the reconstructed
functions are matched very well, i.e. $E(\TwPo)$ is of order
$10^{-10}$, $E(\LiPa)$ is of order $10^{-8}$. Moreover, the anisotropy
measured in terms of $E(\tilde \TwPo)$ was reduces by an order of
magnitude. Plotting $\tilde \TwPo$ only the \PS reconstructions of the
Fontainebleau showed small deviations.

Our explanation for the fact that a slower cooling schedule results in
reconstructions with stronger anisotropy and only poor connectivity is
the artificial anisotropy introduced by reconstructing $\TwPo$ and
$\LiPa$ only in three directions. With increasing number of iterations
the influence of the isotropic, random starting configuration is
decreased while the anisotropic calculation scheme of the two-point
probability functions as described above becomes more significant.
This view agrees with previous work \cite{YT98b,BMHBO99} where 3D
isotropic reconstructions of sandstones using also a fast cooling
schedule were presented.  The poor connectivity of the reconstructions
with slow cooling schedule may be a result of their strong anisotropy.

\begin{figure}

  \begin{center}
    {\epsfig{file=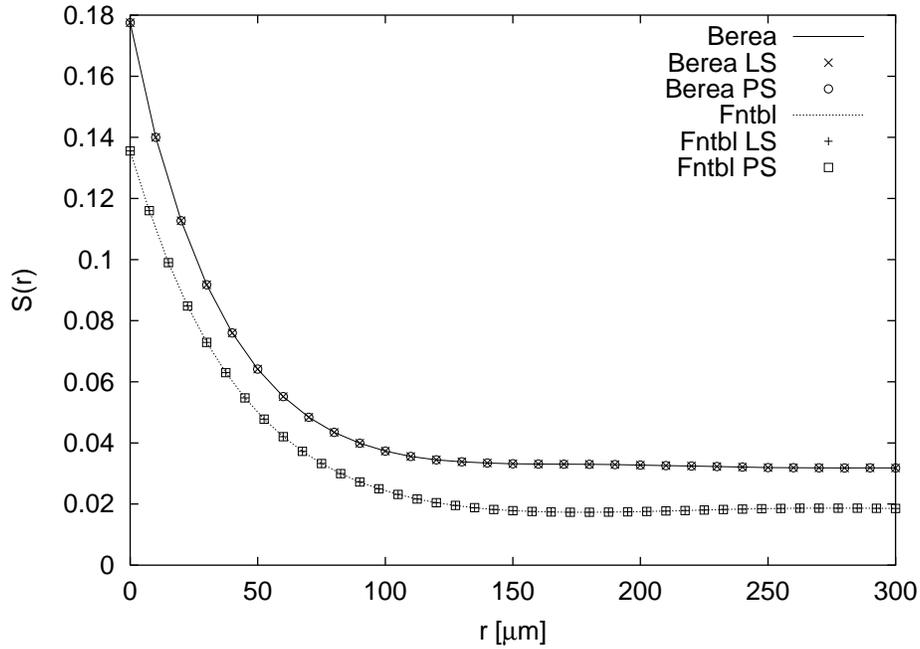,width=.7\linewidth}}
  \end{center}

  \caption
  {Two-point probability functions $\TwPo$ of the Berea (top) and the
  Fontainebleau (bottom) sandstone. The solid lines show the reference
  functions, the points show typical reconstructed functions for a \LS
  and a \PS reconstruction.}
  \label{fig:cf}
\end{figure}

\begin{figure}

  \begin{center}
  {\parbox{\linewidth}{
  \mbox{}\hfill\epsfig{file=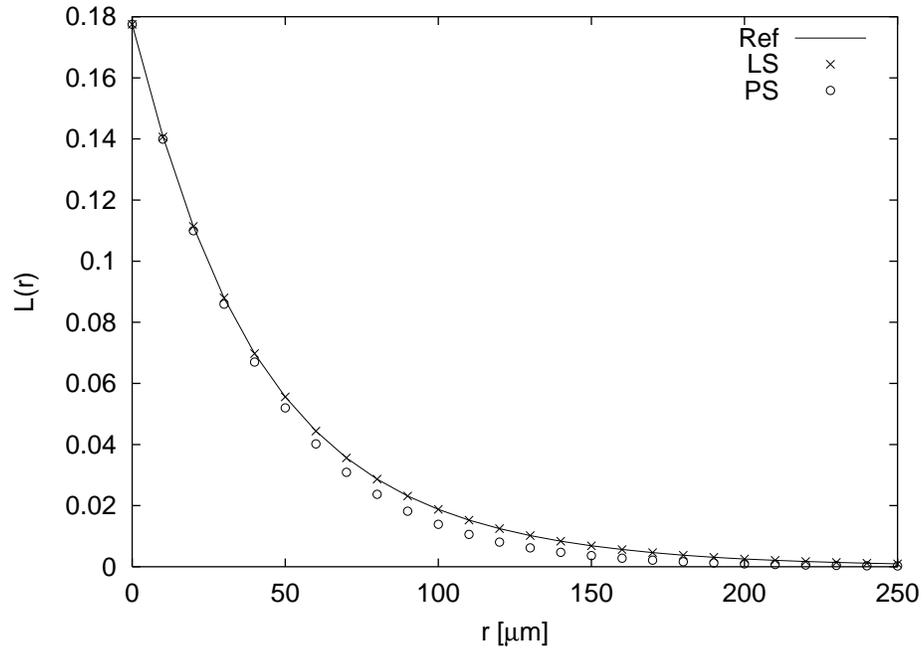,width=.7\linewidth}\hfill
  \vspace*{\baselineskip}\\
  \mbox{}\hfill\epsfig{file=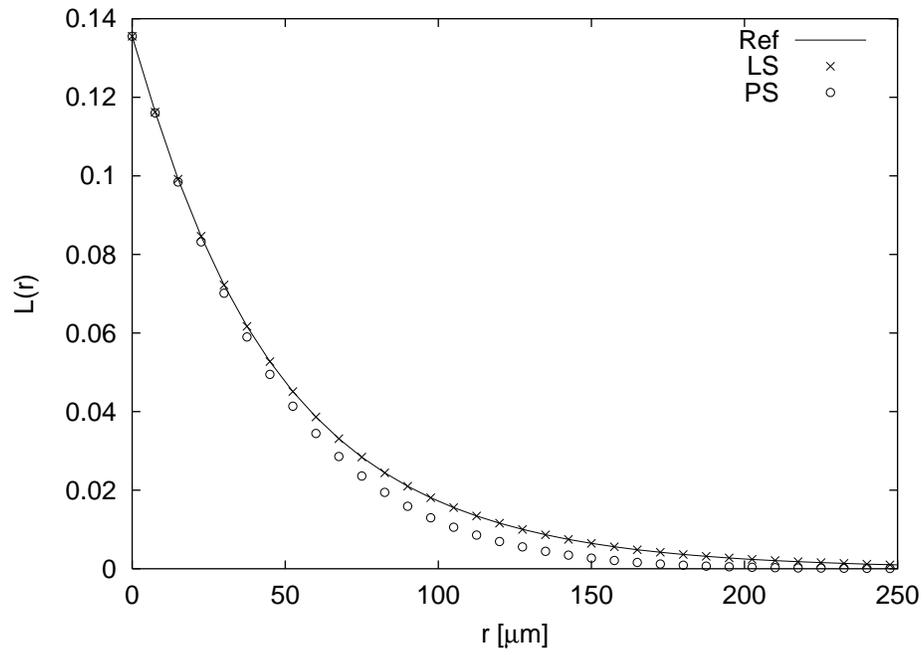,width=.7\linewidth}\hfill\mbox{}
  }}
  \end{center}

  \caption
  {Lineal path functions $\LiPa$ of the Berea (top) and the
  Fontainebleau (bottom) sandstone. The solid lines correspond to the
  reference functions, the points show typical \LS and typical \PS
  reconstructions.}
  \label{fig:lp}
\end{figure}

\begin{figure}

  \begin{center}
  {\parbox{\linewidth}{
  \mbox{}\hfill\epsfig{file=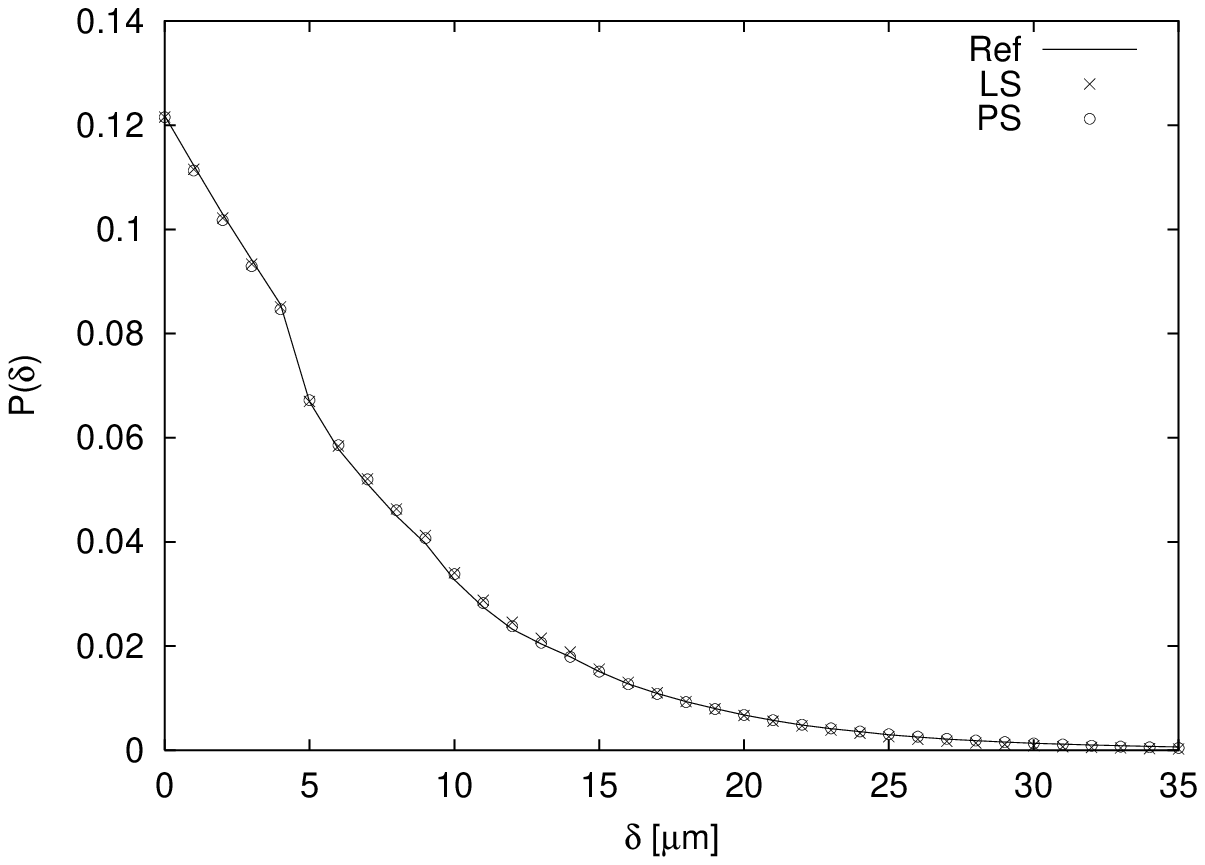,width=.7\linewidth}\hfill
  \vspace*{\baselineskip}\\
  \mbox{}\hfill\epsfig{file=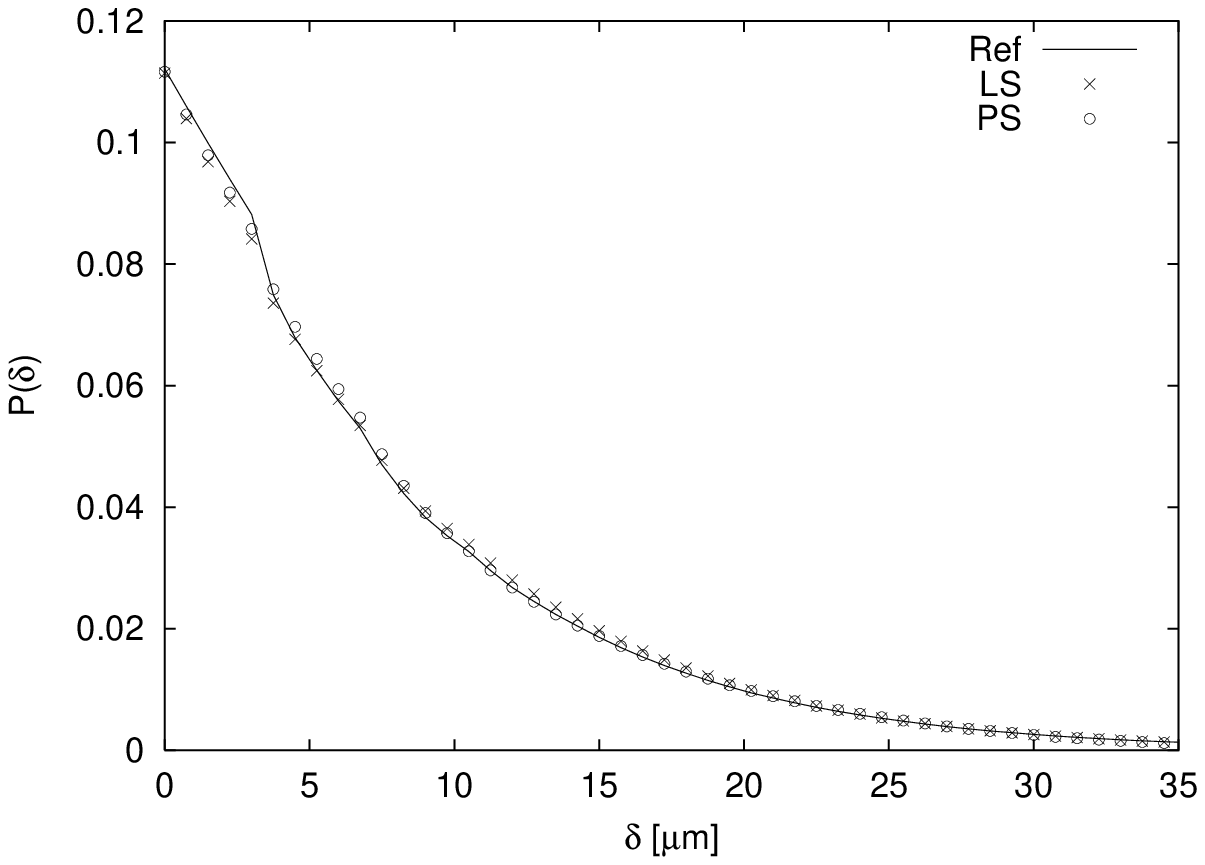,width=.7\linewidth}\hfill\mbox{}
  }}
  \end{center}

  \caption
  {``Pore size'' distribution functions $\PoSi$ of the Berea sandstone (top)
  and the Fontainebleau sandstone (bottom). The solid lines correspond
  to the reference functions, the points show typical \LS and
  typical \PS reconstructions.}
  \label{fig:pdf}
\end{figure}

In Figures \ref{fig:cf}, \ref{fig:lp} and \ref{fig:pdf} the two-point
probability functions $\TwPo$, the lineal path functions $\LiPa$, and
the ``pore size'' distribution function $\PoSi$, respectively, are
plotted for both sandstones using lines and for typical
reconstructions using dots. Here, typical means that the energy given
by Equation (\ref{eq:energy}) of the reconstructed functions is close
to the average value. In the case of the two-point probability
function, the reconstructed functions appear to be indistinguishable
from the reference functions. The same applies to the lineal path
functions measured from the \LS reconstructions. The lineal path
functions of the \PS reconstructions clearly underestimate the
reference functions. Complementary to this, $\PoSi$ is equally matched
by both types of reconstructions as shown in \figref{pdf}.  A
logarithmic plot of $\PoSi$ and $\LiPa$ reveals that the
reconstructions only poorly match the tails of those functions.  This
is due to the extremely small values of $\PoSi(\delta)$ and $\LiPa(r)$
for large $\delta$ and large $r$, respectively.  Furthermore, for the
lineal path function $\LiPa$ this may be a finite size effect because
$\LiPa$ is a long ranged function with $\LiPa(r)>0$ for values of $r$
in the order of a third of the system size.

Looking at Figure \ref{fig:pdf} it seems that for our reconstructions
the lineal path function $\LiPa$ and the two-point probability
function $\TwPo$ incorporate nearly the same information about the
shape of the pores as $\TwPo$ and $\PoSi$ do.  Moreover, as seen from
Figure \ref{fig:lp} the \PS reconstruction lacks information about
long line segments. This may be understood from the fact that $\PoSi$
is a very short ranged function with $\PoSi(\delta)\ne0$ only in a
range smaller than the correlation length. For the Berea sandstone
$\PoSi(\delta)=0$ for $\delta > 60\mum$ and for the Fontainebleau
sandstone $\PoSi(\delta)$=0 for $\delta> 78\mum$. Hence, even though
$\PoSi$ contains full three-dimensional information about spherical
regions in pore space, similar information is already provided by the
two-point probability function $\TwPo$.

Looking at the parameters given in Table \ref{tb:berea} and
\ref{tb:fntbl} the \PS reconstructions match better the specific
surface. This is expected because the specific surface $s$ can be
measured from either $\TwPo$ or $\PoSi$ as seen from Equation
(\ref{eq:ss}) and (\ref{eq:pnod}). However, the value of $s$ measured
from Equation (\ref{eq:pnod}) turns out to be roughly a factor $1.5$
bigger than the value computed from Equation (\ref{eq:ss}). This is
due to the simple surface modeling in the calculation of $\PoSi$ as
discussed above. We believe the kink in $\PoSi(\delta)$ for $\delta =
0.5a$ to be an artifact of the discretization.

In the case of Berea we find the best agreement of the mean pore size
$\bra \delta\ket$ for the \PS reconstructions, while in the case of
the Fontainebleau the match appears to be better for the \LS
reconstruction. The lineal path function seems to be better suited to
describe the long narrow pores of the Fontainebleau sandstone than the
``pore size'' distribution. As already seen from the 2D slices the
appearance of the Berea sandstone is quite similar to the appearance
of the reconstructions. The pores are much more rounded in shape.
Here, the reconstruction is slightly improved with respect to the mean
pore size by incorporating $\PoSi$, which contains information about
spherical regions.

We find analogous results for the mean survival time $\tau$, which is
a diffusive transport property. In fact the mean survival time can be
related to the mean ``pore size'' \cite{TA91}. The mean survival time
is a physical transport property, but unlike the fluid permeability,
it does not capture information about the dynamical connectivity of
the pore space; indeed, neither does the conductivity (or formation
factor). Nonetheless, a cross-property formula relating the fluid
permeability to a combination of the porosity, mean survival time and
formation factor $F$ has been shown to be a highly accurate estimate
of $k$ for sandstones \cite{Sc94}. This cross-property formula was
used to demonstrate that the permeability of another reconstructed
sandstone \cite{YT98b} was in excellent agreement with the exact
Stokes solution determination of the permebaility of the original
sandstone \cite{Sc94}.

We note that a good match of $\tau$ alone may not always indicate a
good match for ${\cal F}$ or $k$.  In fact, combining results from
\cite{WBH99,HWB00} and \cite{BMHBO99} suggest that ${\cal F}$
correlates strongly with the local percolation probability $p$ which
is a measure of the geometrical connectivity.  We find significant
differences between the local percolation probabilities of the real
sandstones and the present reconstructions.

\begin{figure}

  \begin{center}
  {\parbox{\linewidth}{
  \mbox{}\hfill\epsfig{file=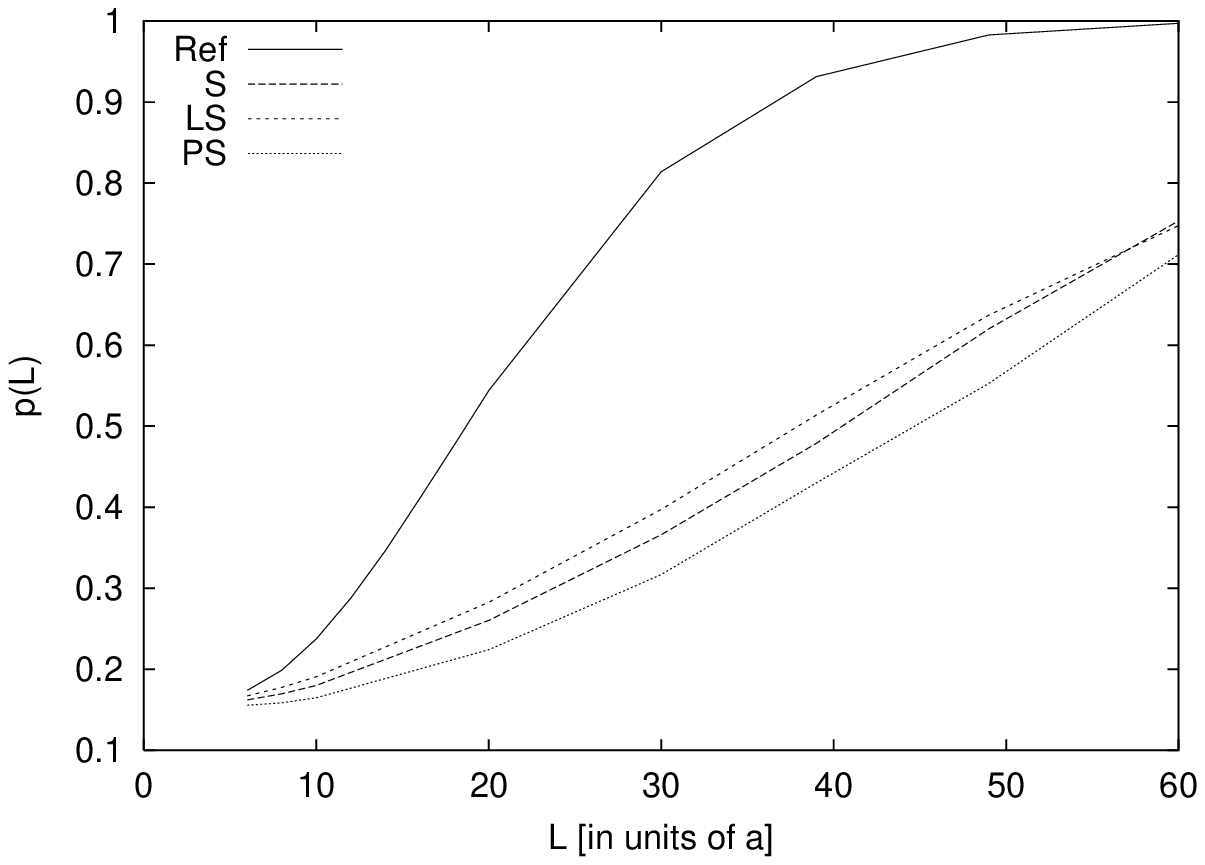,width=.7\linewidth}\hfill
  \vspace*{\baselineskip}\\
  \mbox{}\hfill\epsfig{file=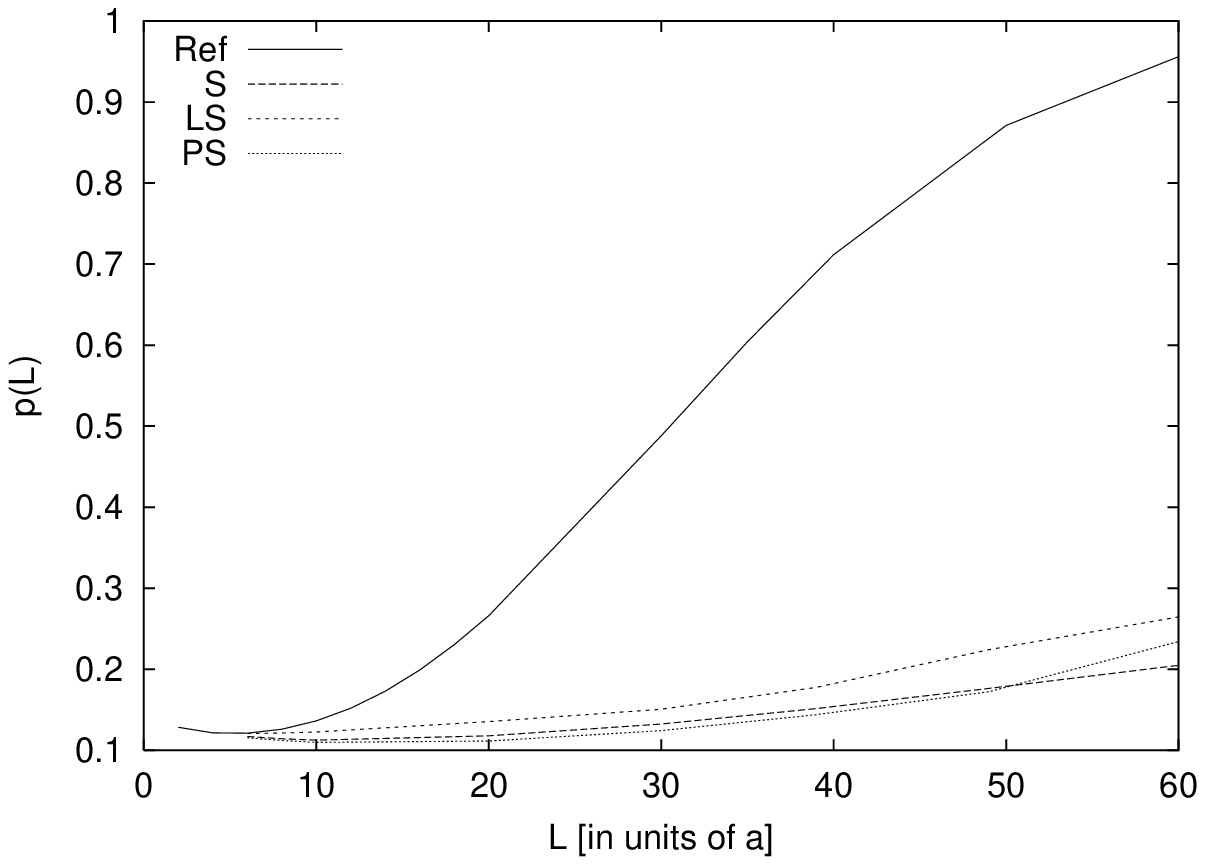,width=.7\linewidth}\hfill\mbox{}
  }}
  \end{center}

  \caption
  {Total fraction of percolating cells $p$ for the Berea sandstone
  (top) and the Fontainebleau sandstone (bottom). The solid lines show
  $p$ measured from the original sandstones. The curves shown for the
  reconstructions are averaged over five configurations each.}
  \label{fig:p}
\end{figure}

\figref{p} shows plots of the local percolation probability $p$ for
the original sandstones, the \LS reconstructions, the \PS
reconstructions, and reconstructions of the two-point probability
function only ({\sffamily S} reconstruction). The pure {\sffamily S}
reconstructions are included here for comparison to previous work
\cite{BMHBO99}.  The curves shown for the reconstructions are averaged
over five configurations each.  The local percolation probability $p$
of the reconstructions lie well below the curves of the original
sandstones.  From this plot it seems, that neither the use of $\LiPa$
nor the use of $\PoSi$ can significantly improve the geometric
connectivity compared to the reconstruction of $\TwPo$ only.  The
differences between the curves of the three reconstructions seems to
be within the range of statistical fluctuations. Nevertheless, a
similar result for the {\sffamily S} reconstruction of the
Fontainebleau sandstone was presented in \cite{BMHBO99} for a larger
sample. Other work \cite{YT98b} showed that a \LS reconstruction
of a different  Fontainebleau sandstone reproduced 
the geometric connectivity well.

In general comparing the \LS reconstructions and the \PS
reconstructions the resulting configurations are quite similar. For
our reconstructions the two-point probability function $\TwPo$ and the
lineal path function $\LiPa$ incorporate nearly the same morphological
information as $\TwPo$ and $\PoSi$ do. Looking at the reconstructions
of the Berea sandstone and the reconstructions of the Fontainebleau
sandstone it appears that the latter one is much more demanding to
reconstruct.  This may be due to its characteristic granular
structure, the narrow pore throats, the lower porosity, and to the
larger sample size of the original sandstone. We also note that the
Berea sample is only $128\times 128\times 128$ resulting in poor
statistical quality.

Our work has shown that simulated annealing provides a flexible and
simple to implement method for reconstructing two-phase random media
and that local porosity theory provides highly sensitive tools for
their comparison and analysis. However, with present computer power
there is still a need to introduce simplifications to reduce the
computation time.  Reconstructing the two-point probability function
$\TwPo$ only in certain directions, may introduce artificially a
strong anisotropy or affect the connectivity. We find, that a fast
cooling schedule can reduce this problem. This implies that the
final configuration is not completely independent from the initial
configuration, and hence the reconstructed microstructure does not
only depend on the reconstructed statistical functions as  would be
desirable.

\section*{Acknowledgments}
C.~M. thanks the Princeton Materials Institute at Princeton University
for their hospitality and C.~Yeong for helpful discussions. C.~M. and
R.~H. gratefully acknowledge financial support by the Deutsche
Forschungsgemeinschaft and the Deutscher Akademischer Austauschdienst.
S.~T. was supported by the Engineering Research Program of the Office
of Basic Energy Sciences at the U.S. Department of Energy under Grant
No. DE-FG02-92ER14275.


\begin{thebibliography}{10}
\bibitem{lpnote}{ The lineal path function is related to the linear
    contact distribution function $\hbox{\sffamily H}_l(r)$ as defined in
    \cite{B:SKM87} with a line segment as the structuring element
    $\BBB$ by the formula $\LiPa(r) = \phi\left(1-\hbox{\sffamily
        H}_l(r)\right)$.  A general contact distribution with
    structuring element $\BBB$ containing the origin may be defined as
    the probability distribution function of the random variable
    $R=\inf\{s:\MMM \cap s\BBB \neq \emptyset\}$ under the condition that
    $R>0$.}
\bibitem{pnote}{ We note that $\hbox{\sffamily F}$ is the same as one minus
    the so-called contact distribution function $\hbox{\sffamily H}_S$,
    $\hbox{\sffamily F}(\delta) = 1 - \hbox{\sffamily H}_S(\delta)$, with a
    sphere of radius $\delta$ as the structuring element $\BBB$
    \cite{B:SKM87}.}
\end{thebibliography}

\onecolumn
\section*{tables}
\begin{table}
\begin{tabular}{l|rrr} 
  & Berea & \LS  & \PS \\
  \hline
  porosity          & $0.1775$ & $0.1775$ & $0.1775$ \\ 
  specific surface   $[{\rm mm}^{-1}]$                     & $13.9$ & $14.7$ & $14.4$ \\
  mean survival time $\tau D$                   $[\mum^2]$ & $100$  &  $89$  & $93$ \\
  mean pore size $\bra \delta \ket$              $[\mum]$  & $6.71$ & $6.52$ & $6.66$ \\
  $f_p$                                          $[\%]$    & $97.16$& $88.76$& $85.49$ \\
  $p(L=60)$                                                & $0.997$& $0.747$& $0.712$ \\               
\end{tabular}
\vspace*{\baselineskip}

\caption{\label{tb:berea} Characteristic quantities of the Berea sandstone, its 
  \LS reconstructions, and its \PS reconstructions. The values given for
  the reconstructions are averaged over five configurations.}
\end{table}

\begin{table}
\begin{tabular}{l|rrr}
  & Fntbl & \LS  & \PS  \\
  \hline
  porosity          & $0.1355$ & $0.1355$ & $0.1355$ \\ 
  specific surface   $[{\rm mm}^{-1}]$                     &$10.0$  & $10.6$   & $10.4$ \\
  mean survival time $\tau D$                    $[\mum^2]$&$134$   &  $129$   & $121$ \\
  mean pore size $\bra \delta \ket$              $[\mum]$  &$7.85$  & $7.88$   & $7.73$   \\
  $f_p$                                          $[\%]$    &$99.35$ & $52.22$  & $51.26$  \\
  $p(L=60)$                                                &$0.956$ & $0.265$  & $0.234$ \\             
\end{tabular}
\vspace*{\baselineskip}

\caption{\label{tb:fntbl} Characteristic quantities of the Fontainebleau sandstone, its 
  \LS reconstructions, and its \PS reconstructions. The values given for
  the reconstructions are averaged over five configurations.}
\end{table}

\newpage
\section*{figures}
\begin{figure}

  \begin{center}
    see figure\_1.jpg
  \end{center}
  
  \caption
  {Two dimensional slices of the Berea sandstone, the Fontainebleau
  sandstone, and two reconstructions of each. The top row shows the
  original sandstones, the middle row the \LS reconstructions, and the
  bottom row the \PS reconstructions with the Berea sandstones on the
  left and the Fontainebleau sandstones on the right side. All slices
  have approximately the average porosity $\phi=0.1775$ for the Berea
  and $\phi=0.1355$ for the Fontainebleau, respectively. }
  \label{fig:slices}
\end{figure}

\end{document}